*Original Contribution*

# Prediction of viral spillover risk based on the mass action principle

**Maryam Golchin[1,2,*], Moreno Di Marco[3], Paul Horwood[2], Dean Paini[2,4], Andrew Hoskins[1,2], R.I. Hickson[1,2,*]**

[1] Commonwealth Scientific and Industrial Research Organisation (CSIRO), Townsville, QLD 4811, Australia; Maryam.Golchin@csiro.au (M.G.); Roslyn.Hickson@csiro.au (R.I.H.); Andrew.Hoskins@csiro.au (A.H.)

[2] College of Public Health Medical and Veterinary Sciences, and Australian Institute of Tropical Health and Medicine, James Cook University, Townsville, QLD 4811, Australia; Paul.Horwood@jcu.edu.au (P.H.)

[3] Department of Biology and Biotechnologies, Sapienza University of Rome, 00185 Roma RM, Italy; Moreno.Dimarco@uniroma1.it (M.D.M.)

[4] CSIRO, Canberra, ACT 2601, Australia; Dean.Paini@csiro.au (D.P.)

* Authors to whom correspondence should be addressed.

**Abstract:** Infectious zoonotic disease emergence, through spillover events, is of global concern and has the potential to cause significant harm to society, as recently demonstrated by COVID-19. More than 70% of the 400 infectious diseases that emerged in the past five decades have a zoonotic origin, including all recent pandemics. There have been several approaches used to predict the risk of spillover through some of the known or suspected infectious disease emergence drivers, largely using correlative approaches. Here, we predict the spatial distribution of spillover risk by approximating general transmission through animal and human interactions. These mass action interactions are approximated through the multiplication of the spatial distribution of zoonotic viral diversity and human population density. Although our results indicate higher risk in regions along the equator and in Southeast Asia where both viral diversity and human population density are high, it should be noted that this is primarily a conceptual exercise. We compared our spillover risk map to key factors, including the model inputs of zoonotic viral diversity estimate map, human population density map, and the spatial distribution of species richness.

Despite the limitations of this approach, this viral spillover map is a step towards developing a more comprehensive spillover risk prediction system to inform global monitoring.

**Keywords:** One Health, pandemic risk, Spillover risk, viral diversity, zoonoses.

**Introduction**

Global efforts to reduce the impacts of emerging zoonotic diseases acknowledge that spillover to humans arises from a complex interplay between humans, animals, and their shared environment. This demonstrates the importance of employing a One Health approach to better understand the underlying drivers and factors of a spillover risk. Indeed, the recent COVID-19 pandemic demonstrated the value of researchers and policymakers collaborating to identify solutions to mitigate the probability of future pandemics. In 2020–21 the Pan-European Commission on Health and Sustainable Development published a report that identified seven objectives with "operationalising the concept of One Health at all levels" being the first one (1).

Spillover can be defined as the cross-species transmission of pathogens, including viruses and bacteria, into susceptible host populations. Spillover can occur between different animal species, such as lumpy skin disease (2) or can transmit from infected animal hosts into human hosts, such as Japanese encephalitis virus (3), Monkeypox (4), and severe acute respiratory syndrome coronavirus (4). Either way, the emergence and spread of infectious diseases have demonstrated a substantial impact on the economy, society and, in the case of zoonotic pathogens, human lives (5,6). Consequently, predicting spillover risk has become a significant focus at a global level.

While known mechanisms and drivers of spillover risk cover the entire One Health spectrum (5), existing studies mostly focus on discovering host-pathogen relationships (7,8) and the risk of pathogens being zoonotic (7–12). For example, the study conducted by Mollentze and Streicker (12) suggests there is no evidence that the taxonomic identity of reservoirs affects the probability that the viruses they harbour are zoonotic. The results of their research indicate that the differences in zoonosis frequency among various animal orders can be understood without needing to propose unique ecological or immunological connections between hosts and viruses. A large number of studies also concentrate on better understanding the drivers and factors that contribute to the spillover risk (8,9,13–15) by implicitly including these drivers in the data sets of statistical or machine learning models. However, a limitation of a purely data-driven approach will inherently be biased by the data available, especially the spatial bias in knowledge of disease emergence events (6,13).

An effective way to mitigate the risk of spillover is by proper management of the human-animal-environment interfaces (8), but data about the interface of spillover risk is scarce (16) (Figure S1 depicts this sparsity for detected pathogen diversity or "richness"). Spillover risk can be identified as the combined probability of successful transmission of a pathogen from an infectious host into a susceptible host (transmission of infection) and the probability of an infection transitioning to a state of disease in the latter (transition to disease) (17). In the case of human spillover, this successful transmission happens when an infectious animal host comes into contact with a susceptible human host, who in turn becomes infected (18,19).

In this study, we focused on a conceptual exercise to explore how well a straightforward and oversimplified mechanism-based approach predicts detected viral spillover risk as a function of pathogen diversity (from wildlife hosts) and human population density. We predicted the spillover risk through principled consideration of the transmission processes. The mass action principle is a basic tenet of epidemiology, relating the number of new infections to the infected population (current number of cases) and susceptible population, requiring contact between those infectious and susceptible. This approach leads to a straightforward model based on zoonotic viral diversity in mammals and birds and population estimates of a susceptible host species (i.e., humans). To explore the validity of our approach, we compared our prediction to existing methods and risks of known drivers such as species richness and human population. This novel but straightforward approach shows promise when compared to known outbreaks, though observation biases in the data are a persistent limitation.

**Material and Methods**

We developed a spatial model of viral spillover risk. We chose the year 2020 as a reference year for human population count data to align with the viral pathogen population data (12) as the proportion of viral viruses in mammal and bird taxonomic orders was generated in 2020.

*Transmission processes – the mass action principle*

Infectious disease transmission is often captured by compartmental mathematical models. The classic compartmental model is the Susceptible-Infectious-Removed (SIR) model, where the population is compartmentalised by these disease statuses. The transition from "susceptible" to "infectious" can only occur when those susceptible come into contact with those infectious. This transition rate is known as the force of infection. Our focus is on the infection term from these compartmental transmission models, which is based on the principle of mass action. The specific infection function depends on whether the transmission is density-dependent (DD) or frequency-dependent (FD) (see, for example, (20)). For a single species, the classic SIR model has a transmission term with the force of infection $\lambda$ and susceptible population $S$ for density-dependent transmission (also known as pseudo-mass action) of the form

$$\lambda S = \beta \times I \times S, \tag{1}$$

and for frequency-dependent transmission (also known as true mass action) of the form

$$\lambda S = \hat{\beta} \times \frac{I}{N} \times S, \tag{2}$$

where $I$ is the infectious population, $N$ is the total population, and $\beta$ and $\hat{\beta}$ are the transmission rates. The transmission rate is composed of the contact rate, which differs by the frequency or density-dependent assumption, and the probability of infection given that contact occurs.

In the DD-based transmission process (1), the number of both susceptible and infectious populations is multiplied, as the contact rate increases with population density. In the FD-based transmission process (2), the number of susceptible and proportion infectious are multiplied, since there is no relation between the contact rate and population density. Since we are focused on cross-species viral transmission where the contact rate likely increases with population densities, all results shown here are for DD transmission with respect to the human population.

The classical SIR model is for a single species, and so if we consider there are $j$ mammals and bird taxonomic order with zoonotic viruses that could transfer into humans, the mass action term would become, for density-dependent transmission,

$$\lambda S = \sum_j \beta_j I_j S. \tag{3}$$

This SIR model approach is not spatially explicit and makes a number of simplifying assumptions including assuming homogeneity within compartments (i.e., all individuals within a compartment are identical). We focus on using the pseudo-mass action principle the classical SIR model is developed from, with further assumptions (outlined in the section "prediction of viral spillover risk" below). For this approach, we require estimates of $I_j$ and $S$, which are described in the next two subsections.

*Zoonotic viral diversity estimate (the $I_j$s)*

The values for $I_j$ was taken directly from a study by Mollentze and Streicker (12). The authors identified viruses likely to be zoonotic, which we then used to construct our viral pool. The authors used literature searches to construct a database of mammals and avian virus-reservoir relationships in conjunction with their histories of human infection. A virus was included in the database when it satisfied three conditions: both humans and multiple independent mammalian or avian reservoir orders maintained the virus with different transmission cycles; the virus infecting humans was confirmed to species level by PCR or sequencing; and a zoonotic maintenance cycle. The final dataset contained eleven animal and bird taxonomic orders (namely Rodentia, Primates, Perissodactyla, Lagomorpha, Diprotodontia, Chiroptera, Cetartiodactyla, Carnivora, Passeriformes, Galliformes, Anseriformes) and the number and proportion of zoonotic virus species associated with each reservoir[1]. This estimate allowed for an implicit consideration of the probability of infection as part of the transmission process outlined in the "Transmission processes" Section.

We accessed mammal and avian species distributional data through the IUCN Red List of Threatened Species (21,22) and filtered the distribution based on the eleven taxonomic orders listed above. The presence feature in IUCN data categorises into Extant, Probably Extant, Possibly Extant, Possibly Extinct Extinct, and Presence Uncertain. We consider the value of 1 for the first three categories as present and removed other categories from our analysis. Furthermore, we filtered data by removing all the samples that were compiled after 2020. We developed the estimates of zoonotic viral diversity (Figure 1) by multiplying the presence data with the proportion of zoonotic viral pathogens and summing up the values for each species at a 1km-by-1km spatial resolution. We note there is an implicit relationship between this estimate and species richness, which we explicitly consider in Figure S4.

*Estimated human population density (the S)*

We used the spatial distribution of human population counts in 2020 from WorldPop (23) at a resolution of 30 arcs (approximately 1km by 1km). This data contains the estimated total number of people per pixel. Details on the approach to calculating the population counts in a pixel can be found in (24–26). We reprojected and resampled this data with sum interpolation in a way that its coordinates reference system, extent, and resolution matched with the detected viral spillover risk data. This data also represented estimated human density per pixel, i.e., the number of people per $km^2$.

---

[1] We highly encourage the interested readers to read through the original paper to understand how data is generated.

*Prediction of relative viral spillover risk*

Our viral spillover risk prediction is based on the pseudo-mass action transmission process, using estimates of viral diversity and human density as described above. That is, to predict a measure of viral spillover risk, we multiplied the zoonotic viral diversity estimate ($\sum_j I_j$), where detection data exists, by human population density ($S$). We then divided the resulting values by the maximum to calculate the relative detected viral spillover risk.

To apply the pseudo-mass action transmission process framework to the spillover risk estimate, we made a number of assumptions. First, we assume a static nature of the spatial transmission, in that we only account for population estimates within a pixel. Second, we do not explicitly account for a relative risk of viral cross-over between species, though they are implicitly accounted for by the proportion of zoonotic viruses (12). Third, we do not take into account the explicit contact rate between species. These latter two assumptions amount to ignoring the transmission rate in the mass action process (the $\beta_j$ and $\widehat{\beta}_j$ parameters in Equations (1) and (2), for $j$ viruses). Due to the lack of data on the animal densities, we have implicitly assumed frequency dependence with respect to the animal populations, and so reference to "frequency" and "density" dependent transmission in our results and discussion is with reference to the human population only. We assume that selected animal orders able to be infected with a known zoonotic virus subsequently pose a risk of infecting humans (implicitly assuming all animals that can be infected are able to infect humans). Since we consider spillover risk to humans of zoonotic viruses, we also assume all humans are susceptible, allowing us to approximate this component of the spillover to humans model by the human population density.

*Comparison between different maps*

To gain a deeper understanding of how our results compared to the state-of-the-art and the underlying factors, we employed a difference-map technique. This involved subtracting our generated maps from other maps to highlight the disparities between them. Specifically, we compared our map of spillover risk (Figure 2) to the zoonotic viral diversity estimate (Figure 1), human population density (Figure S2), and species richness. We further compared our viral spillover risk map with a previous spillover risk prediction (i.e., by Allen *et al.* (13)). To allow direct comparisons, we standardised the values of all the maps by dividing each value by the maximum value of the map, ensuring that all values fall within the range of 0 to 1.

**Results**

*Viral spillover risk predictions*

The distribution of viral diversity is globally uneven with high values across the tropics and lower values in temperate areas, largely following global mammal species richness (Figure 1). Accordingly, our results suggest that spillover risk is elevated in countries across the equator and Southeast Asia where the human population, mammals and avian species richness and zoonotic viral diversity are the highest (Figure 2). Furthermore, we show that spillover risk is lower than human population density in Asia but higher in South America and across the continent of Africa. We interpret this behaviour of the model as the risk of spillover does not rely on human population or species richness, but rather a combination of zoonotic viral diversity and human population.

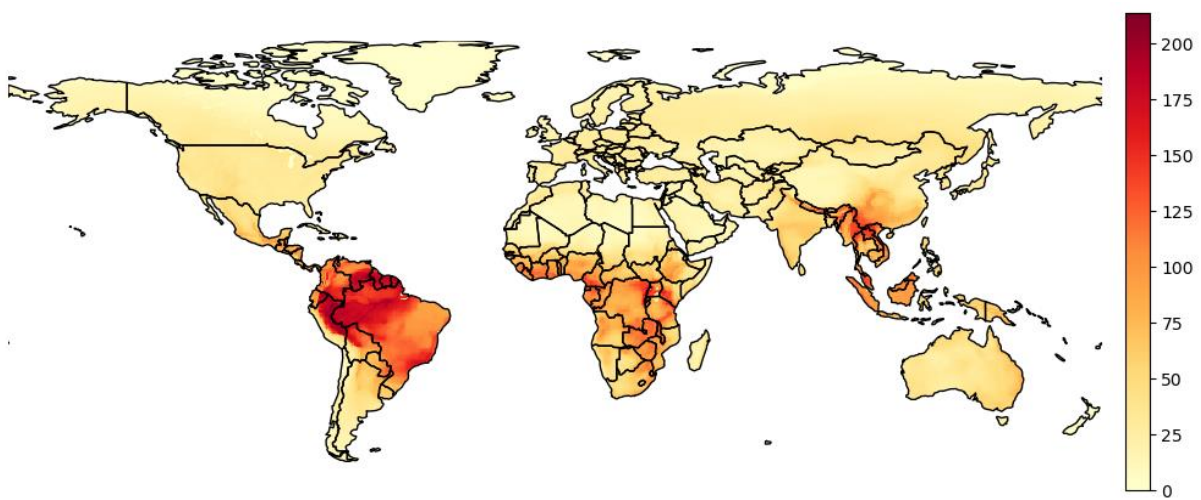

**Figure 1**. The spatial distribution of zoonotic viral diversity, which also represents the frequency-dependent (FD) spillover risk to humans.

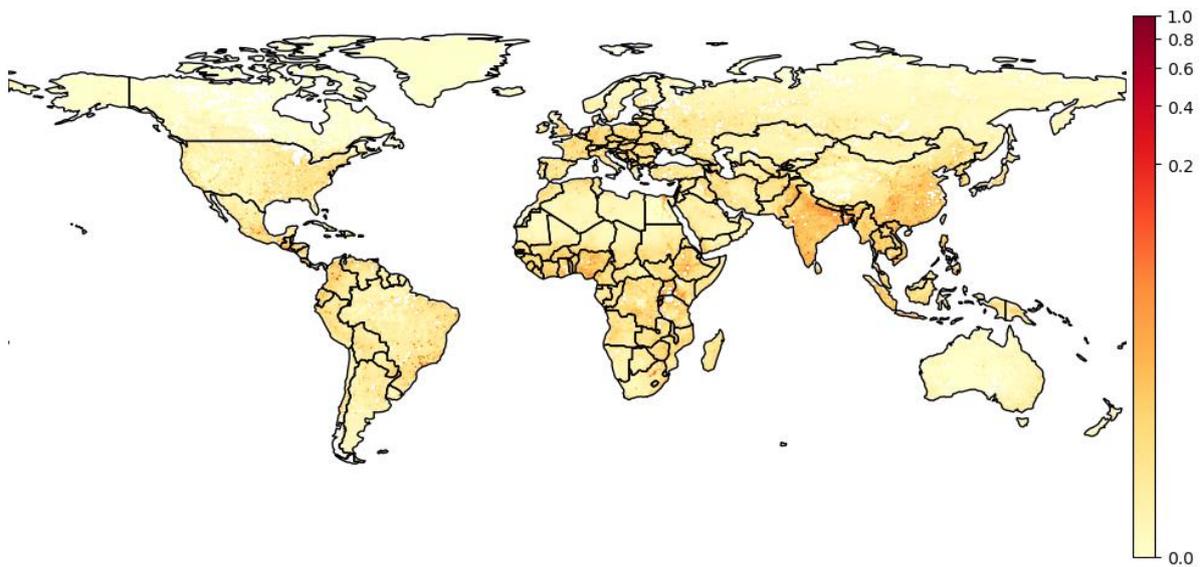

**Figure 2**. Prediction result for relative (density-dependent; DD) spillover risk from animals to humans, based on detected viruses, species distributions, and human population densities. This map is created by linearly mapping a given value to the 0-1 range and then applying a power-law normalisation (i.e., $y = x^\gamma$, where $\gamma$ is the power and set to 0.15) over that range to map colourmaps onto data in non-linear ways. The histogram of the distribution of spillover risk is depicted in Figure S6.

*Comparison between spillover risk maps and other factors*

We compared our prediction of relative viral spillover risk with the zoonotic viral diversity estimate, and underlying factors such as human density estimates, and mammal and bird species richness (27,28) as depicted in Figure 3 and Figure S4. These results suggest that, as expected, areas with high human population density have a relatively higher risk in our spillover risk map, as inherently assumed in many models (13). However, for key regions (see Figure 3 (b)) the potential zoonotic viral diversity acted as an offset, resulting in a relatively lower risk of spillover than a purely human population density-driven risk would suggest.

For example, in Australia or the Amazon rainforest, the spillover risk is lower than expected given the high species richness or zoonotic viral diversity estimate, which indicates species richness or zoonotic viral diversity are not the main drivers of spillover risk. Instead, when we examined human population density, we found no difference between spillover risk and human population density in Australia or the Amazon rainforest, which indicates human population density is the main driver for spillover risk in these areas. In contrast, in Northern India, the spillover risk was higher than expected due to high human population density.

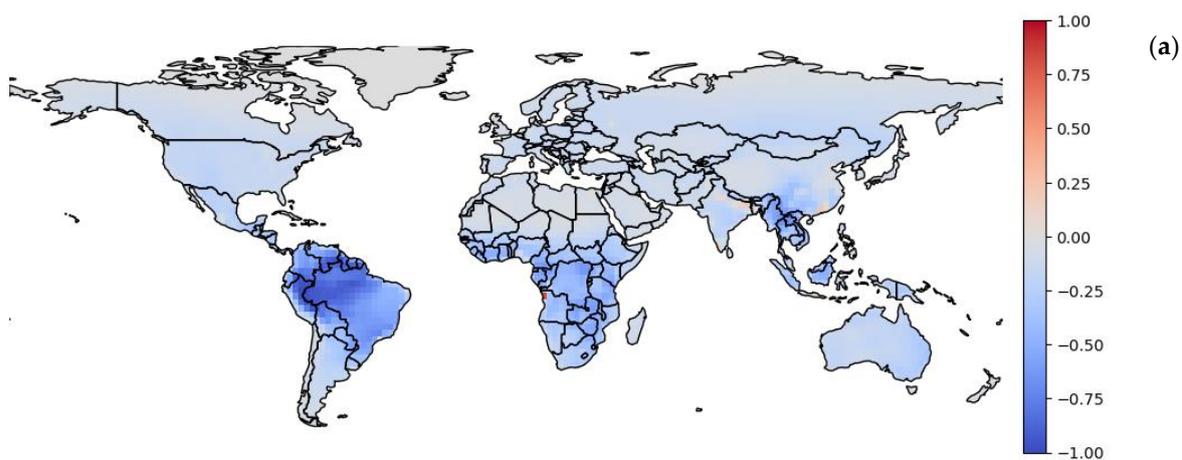

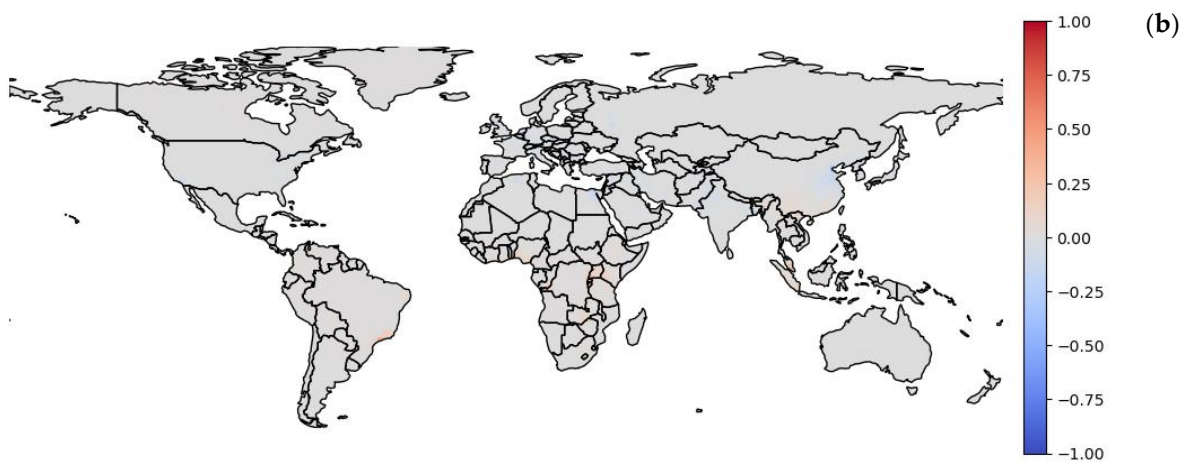

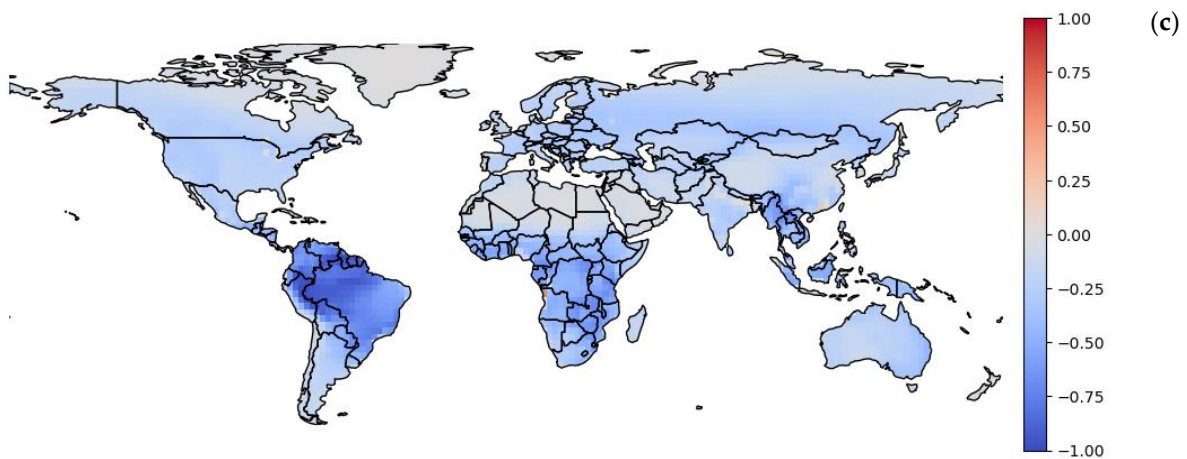

**Figure 3.** The comparison between predicted viral spillover risk and (**a**) zoonotic viral diversity estimate (Figure 1), (**b**) human population density (Figure S2), and (**c**) species richness (Figure S3). We used our predicted spillover risk minus the factor being compared, so positive (or red colour) indicates the predicted viral spillover risk is larger, and conversely negative (or blue colour) values indicate the zoonotic viral diversity estimate, human population density, and species richness are larger, respectively. These maps are created by defining the data range that the colour map covers to -1 and 1.

*Comparison between spillover risk map and previous predictions*

We compared our predicted viral spillover risk map with the result produced by Allen *et al.* (13) in Figure 4. Positive values indicate that our model detected a higher relative viral spillover risk. Figure 4 (a), illustrates the difference-map between our predicted spillover risk and their weighted model with reporting effort. Interestingly, they reported higher risk in countries with higher human activity (the US and countries in Europe). Our results show smaller differences in their refined predictions when they factored out the reporting effort from their result and reweighted their model with the human population (Figure S5). However, the difference in Figure 4 (b) suggests that their results may remain biased towards higher human population densities, regardless of the species richness in those areas (Figure S5).

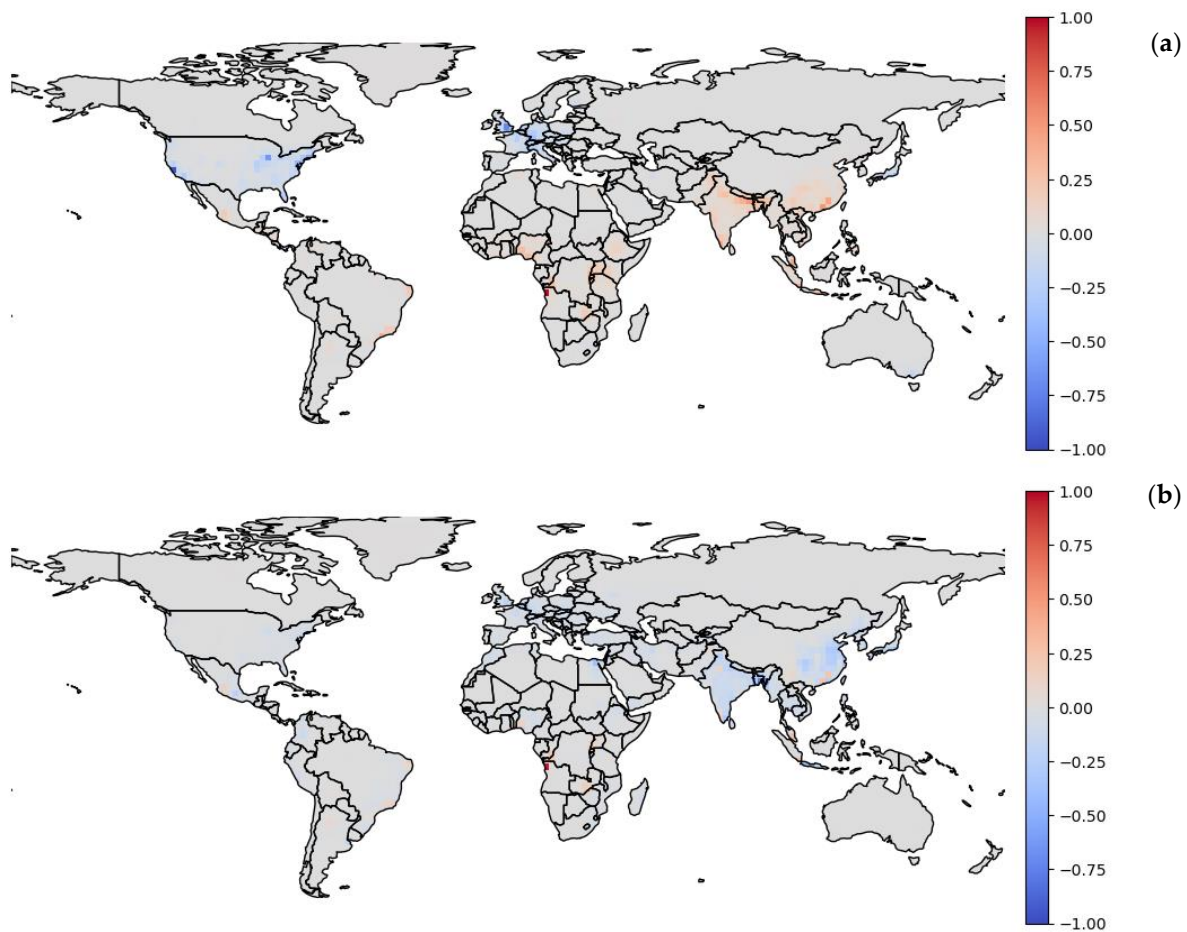

**Figure 4.** The comparison between our model for viral spillover risk and the models by Allen *et al.* (13), (**a**) weighted model output with reporting effort (**b**) weighted model reweighted by population. Positive (red colour) indicates our predicted viral spillover risk is larger, and conversely, negative values (blue colour) indicate the respective Allen *et al.* (13) models are larger. These maps are created by defining the data range that the colour map covers to -1 and 1.

**Discussion**

We developed a spatial model to predict the risk of spillover of viruses known to be able to infect humans, based on the mass action assumption of transmission processes. While simple, this approach showed interesting peculiarities and spatial discrepancies compared to other models not explicitly built on the transmission process (13).

There is a large literature looking at the relationship between biodiversity and the transmission of infectious diseases (29–32). These studies suggest that the risk of new and re-emerging pathogens increases by increasing biodiversity (a type of "amplification"), which increases the risk of spillover. However, higher biodiversity can generate a dilution effect where the presence of many non-competent hosts reduces the overall viral density (33). Additionally, biodiversity loss has a negative impact on human health by increasing the risk of human exposure to both new and re-emerging pathogens (32). Our findings look at the distribution of mammalian and avian taxa groups across the globe and do not directly investigate the effect of biodiversity properties –e.g., species richness—on spillover risk. However, mapping the overlapping geographic ranges of mammal and avian hosts reveals that a higher population of infected host species with viral zoonotic pathogens occur in regions with high species richness, but it does not necessarily mean a higher risk of spillover. Our analyses demonstrate that the difference between spillover risk and zoonotic viral diversity reflects the difference between spillover risk and wildlife species diversity (i.e., richness). Such similarities are expected as species richness is implicitly included in the viral diversity estimate, and the number of zoonotic viruses tends to increase with species richness (12). Moreover, high human population density does not necessarily mean high spillover risk when high human population is separated from high zoonotic viral diversity. The same argument can be made when comparing spillover risk and species diversity. High species diversity does not mean high spillover risk when it is separated from high human population density.

As we used a straightforward modelling framework, there are several limitations to our approach. The work presented here does not investigate the effect of environmental factors in predicting the risk of spillover. Rather it focuses on the principle that pathogen transmission occurs when an infectious host comes into contact with a susceptible host. Figure 3 illustrates the difference between the predicted viral spillover risk and other factors. Specifically, zoonotic viral diversity estimate, which is mathematically equivalent to FD-based spillover risk with respect to animals for this subset of viruses (Figure 3 (a)) and human population density (Figure 3 (b)).

We made a number of assumptions in building this model. The results shown throughout assume a density-dependent transmission process, with respect to the human population. As expected, a DD-based model is more sensitive to population densities and predicts a higher risk when we have a higher human population density. A refinement of the assumption that species crossover is equally likely and only depends on relative population pressures should also be explored before using this prediction to inform responses. This means our raw risk prediction represents an upper bound, but our relative risk would change in areas where humans have higher or lower susceptibility. Additionally, the most likely transmission process for spillover is density-dependent with respect to both animal and human populations, though here we implicitly assumed a frequency-dependent process with respect to animals due to a lack of animal density estimates. This is a useful step towards a principled estimate, but the spatial patterns would likely substantially change with higher relative risk in areas with denser animal populations (further removing the bias towards human population density).

Allen *et al.* (13) selected a refined set of spatial predictors for their relevance to a priori hypotheses on plausible mechanisms underlying zoonotic disease emergence, including proxies for human activity (reporting effort), environmental factors, and the zoonotic pathogen pool from which novel diseases could emerge. Their results suggest that emerging infectious diseases of wildlife origin are more likely to occur in regions with higher human populations, greater wildlife diversity, and greater levels of land-use change. Given the required contact across the human-animal interface is, on average, dependent on the respective population densities, we think the density-dependent approach is more appropriate. Our difference maps further suggest that even with the correction for human population density, there is still a source of bias in the Allen *et al.* prediction (Figure 4 (a)).

A major limitation is that our estimates of relative spillover risk have not been validated against an independent data source.

Conclusion

In this study, we undertook a conceptual approach to explore spillover risk prediction based on the mass-action principle underlying transmission processes, using data for zoonotic viruses, relevant animal population distributions, and human density. We subsequently predicted both a global map of viral diversity and viral spillover risk for the subset of viruses known to be zoonotic. We implicitly used a One Health approach by considering humans, animals, and the environment in our modelling. We suggest that Figure 3 (b) might be useful for the planning of zoonotic risk monitoring in areas at high potential risk of spillover, where early warning signals would be especially valuable. This is an important step towards the prevention of zoonotic epidemics and pandemics (34), albeit an often under-resourced one (35). Although this simple approach seems promising, there are multiple avenues for further improvements, including ground truthing or validating this approach with independent data. Future work to improve our prediction also includes explicitly capturing the spatiotemporal interactions between key One Health systems and encoding these into transmission pathways to predict aggregated global spillover risks. We believe our method could work for all pathogens, providing there is sufficient knowledge of their wildlife host species. We also believe our approach has the potential to be used to forecast spillover risk under future scenarios (9), to help form the basis of a simulation system that can inform policy decisions.

Author Contributions:

methodology R.I.H., M.G., and A.H.; software, M.G. and A.H.; validation, M.G. and R.I.H.; formal analysis, M.G., R.I.H., and A.H.; investigation, A.H., R.I.H., M.G.; data curation, M.G., and A.H.; writing—original draft preparation, M.G., and R.I.H.; writing—review and editing, R.I.H., M.D.M., D.P., A.H., P.H., and M.G.; visualization, M.G.; supervision, R.I.H.; project administration, R.I.H., A.H., and M.G. All authors have read and agreed to the published version of the manuscript.

Data availability Statement

Links to the code and data to generate the results of this paper will be provided on request.

Declaration of Interest

None.

**References**

1. Forman R, Azzopardi-Muscat N, Kirkby V, Lessof S, Nathan NL, Pastorino G, et al. Drawing light from the pandemic: Rethinking strategies for health policy and beyond. Health Policy. 2022 Jan 1;126(1):1–6.

2. Coetzer J a. W, Tustin RC. Infectious diseases of livestock. Volume Three. Infectious diseases of livestock Volume Three [Internet]. 2004 [cited 2022 Aug 30];(Ed.2). Available from: https://www.cabdirect.org/cabdirect/abstract/20053143412

3. Ghosh D, Basu A. Japanese Encephalitis—A Pathological and Clinical Perspective. PLoS Negl Trop Dis. 2009 Sep 29;3(9):e437.

4. Osorio JE, Yuill TM. Zoonoses. In: Mahy BWJ, Van Regenmortel MHV, editors. Encyclopedia of Virology (Third Edition) [Internet]. Oxford: Academic Press; 2008 [cited 2022 Aug 30]. p. 485–95. Available from: https://www.sciencedirect.com/science/article/pii/B9780123744104005367

5. Randolph DG, Refisch J, MacMillan S, Wright CY, Bett B. Preventing the next pandemic - Zoonotic diseases and how to break the chain of transmission. UNEP; 2020 p. 82. (UNEP's Frontiers Report Series).

6. Jones KE, Patel NG, Levy MA, Storeygard A, Balk D, Gittleman JL, et al. Global trends in emerging infectious diseases. Nature. 2008 Feb;451(7181):990–3.

7. Becker DJ, Albery GF, Sjodin AR, Poisot T, Bergner LM, Chen B, et al. Optimising predictive models to prioritise viral discovery in zoonotic reservoirs. The Lancet Microbe [Internet]. 2022 Jan 10 [cited 2022 Feb 1]; Available from: https://www.sciencedirect.com/science/article/pii/S2666524721002457

8. Roberts M, Dobson A, Restif O, Wells K. Challenges in modelling the dynamics of infectious diseases at the wildlife–human interface. Epidemics. 2021 Dec 1;37:100523.

9. Carlson CJ, Albery GF, Merow C, Trisos CH, Zipfel CM, Eskew EA, et al. Climate change increases cross-species viral transmission risk. Nature [Internet]. 2022 Apr 28 [cited 2022 Apr 29]; Available from: https://www.nature.com/articles/s41586-022-04788-w

10. Grange ZL, Goldstein T, Johnson CK, Anthony S, Gilardi K, Daszak P, et al. Ranking the risk of animal-to-human spillover for newly discovered viruses. Proc Natl Acad Sci USA. 2021 Apr 13;118(15):e2002324118.

11. Singh BB, Ward MP, Dhand NK. Inherent virus characteristics and host range drive the zoonotic and emerging potential of viruses. Transboundary and Emerging Diseases [Internet]. 2021 [cited 2022 May 24];n/a(n/a). Available from: https://onlinelibrary.wiley.com/doi/abs/10.1111/tbed.14361

FOR PEER REVIEW                                                                                                                                                17 of 1812. Mollentze N, Streicker DG. Viral zoonotic risk is homogenous among taxonomic orders of mammalian and avian reservoir hosts. Proc Natl Acad Sci USA. 2020 Apr 28;117(17):9423–30.

13. Allen T, Murray KA, Zambrana-Torrelio C, Morse SS, Rondinini C, Di Marco M, et al. Global hotspots and correlates of emerging zoonotic diseases. Nat Commun. 2017 Dec;8(1):1124.

14. Saylors K, Wolking DJ, Hagan E, Martinez S, Francisco L, Euren J, et al. Socializing One Health: an innovative strategy to investigate social and behavioral risks of emerging viral threats. One Health Outlook. 2021 May 14;3(1):11.

15. Napolitano Ferreira M, Ellio W, Golden Kroner R, Kinnaird MF, Prist PR, Valdujo P, et al. Drivers and causes of zoonotic diseases: an overview. PARKS. 2021 Mar 11;(27):15–24.

16. Meurens F, Dunoyer C, Fourichon C, Gerdts V, Haddad N, Kortekaas J, et al. Animal board invited review: Risks of zoonotic disease emergence at the interface of wildlife and livestock systems. Animal. 2021 Jun;15(6):100241.

17. Power AG, Mitchell CE. Pathogen Spillover in Disease Epidemics. The American Naturalist. 2004 Nov;164(S5):S79–89.

18. Ellwanger JH, Chies JAB. Zoonotic spillover: Understanding basic aspects for better prevention. Genet Mol Biol. 2021;44(1 Suppl 1):e20200355.

19. Alexander KA, Carlson CJ, Lewis BL, Getz WM, Marathe MV, Eubank SG, et al. The Ecology of Pathogen Spillover and Disease Emergence at the Human-Wildlife-Environment Interface. The Connections Between Ecology and Infectious Disease. 2018 Apr 28;5:267–98.

20. Cohen T, White P. Transmission-dynamic models of infectious diseases. In: Abubakar I, Stagg HR, Cohen T, Rodrigues LC, editors. Infectious Disease Epidemiology [Internet]. Oxford University Press; 2016 [cited 2022 Aug 30]. p. 0. Available from: https://doi.org/10.1093/med/9780198719830.003.0016

21. IUCN [Internet]. 2022 [cited 2023 Sep 1]. Available from: https://www.iucnredlist.org

22. BirdLife International and Handbook of the Birds of the World [Internet]. 2022 [cited 2023 Sep 1]. Available from: http://datazone.birdlife.org/species/requestdis

23. WorldPop. Global 1km Population [Internet]. University of Southampton; 2018 [cited 2022 Jul 25]. Available from: https://www.worldpop.org/doi/10.5258/SOTON/WP00647

24. Gaughan AE, Stevens FR, Huang Z, Nieves JJ, Sorichetta A, Lai S, et al. Spatiotemporal patterns of population in mainland China, 1990 to 2010. Sci Data. 2016 Feb 16;3(1):160005.

25. Sorichetta A, Hornby GM, Stevens FR, Gaughan AE, Linard C, Tatem AJ. High-resolution gridded population datasets for Latin America and the Caribbean in 2010, 2015, and 2020. Sci Data. 2015 Sep 1;2(1):150045.

26. Stevens FR, Gaughan AE, Linard C, Tatem AJ. Disaggregating Census Data for Population Mapping Using Random Forests with Remotely-Sensed and Ancillary Data. PLOS ONE. 2015 Feb 17;10(2):e0107042.

27. Jenkins CN, Pimm SL, Joppa LN. Global patterns of terrestrial vertebrate diversity and conservation. Proceedings of the National Academy of Sciences. 2013 Jul 9;110(28):E2602–10.

28. Pimm SL, Jenkins CN, Abell R, Brooks TM, Gittleman JL, Joppa LN, et al. The biodiversity of species and their rates of extinction, distribution, and protection. Science. 2014 May 30;344(6187):1246752.

29. Keesing F, Ostfeld RS. Impacts of biodiversity and biodiversity loss on zoonotic diseases. PNAS [Internet]. 2021 Apr 27 [cited 2021 Nov 21];118(17). Available from: https://www.pnas.org/content/118/17/e2023540118

30. Rahman MdT, Sobur MdA, Islam MdS, Ievy S, Hossain MdJ, El Zowalaty ME, et al. Zoonotic Diseases: Etiology, Impact, and Control. Microorganisms. 2020 Sep 12;8(9):1405.

31. Luis AD, Kuenzi AJ, Mills JN. Species diversity concurrently dilutes and amplifies transmission in a zoonotic host–pathogen system through competing mechanisms. Proceedings of the National Academy of Sciences. 2018 Jul 31;115(31):7979–84.

32. Keesing F, Belden LK, Daszak P, Dobson A, Harvell CD, Holt RD, et al. Impacts of biodiversity on the emergence and transmission of infectious diseases. Nature. 2010 Dec;468(7324):647–52.

33. Johnson PTJ, Thieltges DW. Diversity, decoys and the dilution effect: how ecological communities affect disease risk. Journal of Experimental Biology. 2010 Mar 15;213(6):961–70.

34. Di Marco M, Baker ML, Daszak P, De Barro P, Eskew EA, Godde CM, et al. Sustainable development must account for pandemic risk. Proceedings of the National Academy of Sciences. 2020 Feb 25;117(8):3888–92.

35. Bernstein AS, Ando AW, Loch-Temzelides T, Vale MM, Li BV, Li H, et al. The costs and benefits of primary prevention of zoonotic pandemics. Science Advances. 2022 Feb 4;8(5):eabl4183.